\documentclass[fleqn,twoside]{article}
\usepackage{espcrc2}

\usepackage{graphicx}
\usepackage[figuresright]{rotating}

\newcommand{\AmS}{{\protect\the\textfont2
  A\kern-.1667em\lower.5ex\hbox{M}\kern-.125emS}}

\title{Spin excitations in stripe-ordered La$_{2-x}$Sr$_x$NiO$_4$ ($x=0.275$ and 1/3)}

\author{A.T. Boothroyd\address[Oxford]{Department of Physics,
        University of Oxford,
        Oxford, OX1 3PU, United Kingdom}$^,$%
        \thanks{Corresponding author. Tel.: +44 1865 272376; fax: +44 1865 272400;
        {\it E-mail address:} a.boothroyd1@physics.ox.ac.uk.},
        P.G. Freeman\addressmark,
        D. Prabhakaran\addressmark,
        H. Woo\address[ISIS]{ISIS Facility, Rutherford Appleton Laboratory,
        Chilton, Didcot, OX11 0QX, United Kingdom}$^,$\address[BNL]{Department of Physics, Brookhaven National Laboratory,
        Upton, New York 11973, USA},
        K. Nakajima\address{Neutron Scattering Laboratory,
        ISSP, University of Tokyo, Tokai, Ibaraki, Japan},
        J.M. Tranquada\addressmark[BNL],
        K. Yamada\address{Institute for Chemical Research, Kyoto University, Gokasho,
Uji 610--0011, Japan},
        and
        C.D. Frost\addressmark[ISIS]}

\begin{document}

\begin{abstract}
We report neutron scattering measurements of the spectrum of
magnetic excitations in the stripe-ordered phase of
La$_{2-x}$Sr$_x$NiO$_4$ ($x=0.275$ and 1/3). The propagating spin
excitations follow a similar dispersion relation for the two
compositions, but the line widths are broader for $x=0.275$ than
for $x=1/3$. \vspace{1pc}
\end{abstract}

\maketitle


It is now well established that charge carriers doped into
antiferromagnets tend to segregate into parallel domain walls
separating regions with antiferromagnetic (AF) order of the host
spins. This so-called stripe ordering is observed in members of
the cuprate family of superconductors
\cite{Tranquada-Nature-1995},
making it an exciting field for basic research.

The stripe modulation of spin and charge has been characterized
extensively in the layered La$_{2-x}$Sr$_x$NiO$_{4+y}$ system
\cite{Chen-PRL-1993,Tranquada-PRL-1994,Yoshizawa-PRB-2000}, where
$n_{\rm h} = x+2y$ is the level of hole doping. Interest is now
turning to stripe dynamics. Recently, we reported two separate
neutron inelastic scattering studies of the magnetic excitations
in the stripe phase of La$_{2-x}$Sr$_x$NiO$_4$ with $x\simeq 1/3$
\cite{Boothroyd-PRB-2003,Bourges-CM/0203187}. In both cases we
observed propagating spin-wave modes throughout the Brillouin
zone, with a maximum energy of 80--85\, meV.

The case of $x=1/3$ is special in that the periodicities of the
charge and magnetic superlattices are the same and are
commensurate with the host lattice. This leads to a very stable
stripe order, with a relatively high ordering temperature of
200\,K and an in-plane correlation length of several hundred $\rm
\AA$ \cite{Yoshizawa-PRB-2000}.  Here we compare the magnetic
excitation spectrum in crystals with $x=0.275$ and $x=1/3$. These
crystals have similar stripe correlation lengths, but at $x=0.275$
the spin and charge periodicities are not the same and are not
commensurate with the host lattice.

Neutron inelastic scattering measurements were made on the MAPS
chopper spectrometer at the ISIS Facility. The single-crystal
samples were mounted in a closed-cycle refrigerator and aligned
with the $c$ axis parallel to the incident beam direction.
Scattered neutrons were recorded in large banks of
position-sensitive detectors. We analyzed the spectra by making a
series of constant-energy slices and projecting the intensity on
to the $(h, k, 0)$ reciprocal lattice plane. The $c$ component of
the scattering vector is unimportant as there are no measurable
spin correlations along the $c$ axis in the range of energies
probed.

The diagonal stripe pattern in La$_{2-x}$Sr$_x$NiO$_4$ is twinned,
with stripes either parallel to the $[1,-1,0]$ or $[1,1,0]$
directions of the tetragonal lattice (cell parameter $a=3.8\,{\rm
\AA}$). These are described by magnetic ordering wave vectors
$(\frac{1}{2},\frac{1}{2},l) \pm (\delta, \delta, 0)$ and
$(\frac{1}{2},\frac{1}{2},l) \pm (\delta, -\delta, 0)$,
respectively, where $\delta \simeq n_{\rm h}/2$. In our samples we
observed an equal population of the twin domains.

\begin{figure}[htb]
\includegraphics*{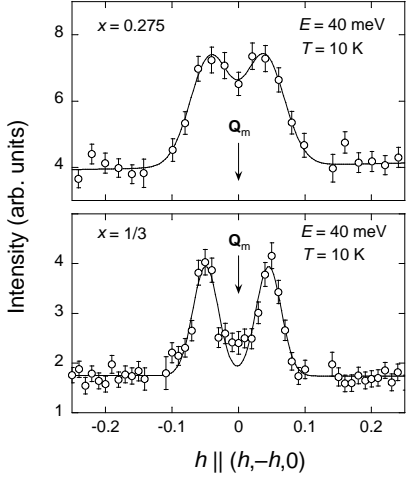}
\caption{\small Constant-energy scans at $E=40$\,meV through the
magnetic zone centres in stripe-ordered La$_{2-x}$Sr$_x$NiO$_4$
($x=0.275$ and 1/3). The scans run parallel to the stripe
direction. The lines are fits to Gaussian peak profiles.}
\label{fig1}
\end{figure}

Figure \ref{fig1} shows scans at an energy of 40\,meV through the
magnetic zone centres {\bf Q}$_{\rm m} = (0.35, 0.35, l)$ for
$x=0.275$ and {\bf Q}$_{\rm m} = (1/3, 1/3, l)$ for $x=1/3$, with
$l\simeq 5$ in both cases. The peaks correspond to spin waves
propagating parallel to the stripe direction. The positions of the
spin-wave peaks are similar for both crystals, but the peaks are
much broader for $x=0.275$ and only just resolved. At lower
energies the widths of the $x=0.275$ spin-wave peaks decrease, but
are always greater than for $x=1/3$.

\begin{figure}[htb]
\includegraphics*{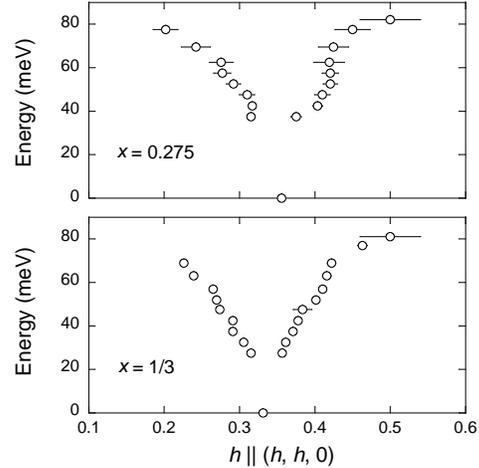}
\caption{\small Spin wave dispersion perpendicular to the stripe
direction in La$_{2-x}$Sr$_x$NiO$_4$. The points at zero energy
are from the magnetic order Bragg peaks.} \label{fig2}
\end{figure}

By performing Gaussian fits to scans with different energies we
determined the two-dimensional spin-wave dispersion. The
dispersion perpendicular to the stripe direction is shown in Fig.
\ref{fig2}. Reliable values for the peak centres could only be
obtained for energies above $\sim30$\,meV. In this energy range
the dispersion for the two crystals is similar, and consistent
with that found in Refs.
\cite{Boothroyd-PRB-2003,Bourges-CM/0203187}. Some differences
were observed below 30\,meV, and these will be discussed
elsewhere.

Perhaps the most interesting result from this study is the fact
that the spin excitations of the $x=0.275$ phase possess a
significant intrinsic width which increases with energy, whereas
those of $x=1/3$ do not. Remember that both compositions exhibit
very long range stripe order.

It is possible that the robustness of the $x=1/3$ stripes is due
to a strong pinning of the commensurate stripes to the lattice. On
the other hand, a clear anomaly has been found near 15--20\,meV in
the spin excitation spectrum of $x=1/3$ \cite{Boothroyd-PRB-2003}.
This may be due to some kind of resonant spin-charge coupling, and
investigation of this energy range in the $x=0.275$ sample will be
of interest.

\end{document}